\documentclass[a4paper,11pt]{article}
\pdfoutput=1

\usepackage[pdftex]{hyperref}
\usepackage{amsmath}
\usepackage{amssymb}
\usepackage{fullpage}
\usepackage{cancel}
\usepackage{float}
\usepackage[affil-it]{authblk}
\usepackage[hang,flushmargin]{footmisc}
\usepackage{color}

\newcommand{\ar}{$a_r^{\! (1)}\,$}
\newcommand{\ath}{$a_3^{\! (1)}\,$}
\newcommand{\athf}{$a_3^{\! (1)}$}
\newcommand{\arf}{$a_3^{\! (1)}$}

\numberwithin{equation}{section}

\hypersetup{ pdfborder = {0 0 0}}

\title{Defect fusing rules in affine Toda field theory}
\author{Craig Robertson\footnote{\href{mailto:craig.robertson@durham.ac.uk}{craig.robertson@durham.ac.uk}}}
\date{}

\begin{document}

\begin{titlepage}

\renewcommand{\thefootnote}{\alph{footnote}}
\clearpage\maketitle
\thispagestyle{empty}
\renewcommand{\thefootnote}{\arabic{footnote}}

\begin{abstract}
\noindent
The concept of fusing rules for defects in the \ar affine Toda field theories is introduced in the classical and quantum contexts. This idea is employed in finding a new transmission matrix for the \ath theory which obeys the triangle relations and the crossing-unitarity relations.
\end{abstract}
\end{titlepage}

\newpage

\section{Introduction}

Affine Toda field theory (ATFT) is a subject which has a long history \cite{MOP,Wils,OT1,OT2} but continues to be of interest today with much of the current interest stemming from the discovery of integrable defects. Defects were initially discovered in the $a_1$ (sine-Gordon) model \cite{KL,BCZ03} and it is this model has dominated the literature on affine Toda defects \cite{KL,BCZ03,BCZ05,HK,BS,Nem,CZ10a,AAGZ,AD}. Defects have also been found in the other \ar models \cite{BCZ04} (and those ATFTs that can be found by folding \ar \cite{CZ09b,Rob}) but the situation there is much less complete, while for the other simply laced theories there aren't even any known integrable defects. The eventual goal of this line of enquiry is to find and investigate the properties of all of the possible affine Toda defects.
\\
\\
Fusing rules (or fusion rules) allow the properties of all of the fundamental particles of the theory to be determined in terms of a minimal set of `basic' particles. In ATFT fusing rules have long been known for the fundamental excitations \cite{Dor90,Dor91,FO} and also for the solitons \cite{Holl92,OTU93,Hall}. In finding the soliton scattering matrices ($S$-matrices) or the transmission matrices for solitons through defects ($T$-matrices) one typically only needs to consider what happens to the basic solitons, as the rest follows by the fusing rules.
\\
\\
There are reasons to treat defects as being particles on an equal footing with solitons. One reason is that defects possess their own energy and momentum \cite{CZ09a}; while another reason is that $S$-matrices have been found embedded in $T$-matrices \cite{CZ10a,CZ10b}, suggesting that certain defect configurations can mimic solitons. By treating defects as particles there ought to exist defect fusing rules, which would go a long way towards systematising the study of defects in ATFT. Defect fusing rules have previously been considered in conformal \cite{PZ} and perturbed conformal \cite{Run} field theories, but not the affine Toda theories. Although $a_1$ does not possess fusing rules; the other \ar theories do, so provide a suitable arena for the study of defect fusing.
\\
\\
In this paper the idea of defect fusing rules in \ar is introduced with the premise that \ar possesses $r$ species of fundamental defect in analogy to how it possesses $r$ species of fundamental soliton. In fact, there are remarkable similarities between soliton and defect fusing rules so analogy to soliton fusing is made throughout. The idea is explained at the classical and quantum levels before it is applied to find a new transmission matrix for \ath - the lowest rank ATFT for which there should be a new fundamental defect not previously considered in \cite{CZ07,CZ09a}. This paper does not solve the general \ar defect fusing equation \eqref{genfuse}, but finding the general solution for the defect fusing couplings would allow for the systematic study of \ar defects.
\\
\\
In order to set notation, a short summary of the the relevant concepts in ATFT with defects is given below.
\\
\\
A 1+1 dimensional affine Toda field theory can be associated to each affine Dynkin diagram \cite{OT2}. The bulk\footnote{The `bulk' is the region away from any defects or boundaries, so the bulk Lagrangian describes the theory in the absence of defects and boundaries.} ATFT Lagrangian is
\begin{align}
\mathcal{L} = \frac{1}{2}\dot{u} \cdot \dot{u} - \frac{1}{2} u' \cdot u' - U(u) \label{bulk}
\end{align}
with potential
\begin{align}
U(u) = \frac{m^2}{\beta^2}\sum_{j=0}^r n_j \left( e^{\beta \alpha_j \cdot u} - 1 \right) \; . \label{pot}
\end{align}
\\
In \eqref{bulk} and \eqref{pot} the field $u$ is an $r$ component vector living in the root space in question with $\{ \alpha_i \}$ the positive simple roots and $\alpha_0 = - \sum_{j=1}^r n_j \alpha_j$ the lowest root in the root space. By convention, $n_0 = 1$, so the other marks $\{ n_i \}$ are a characteristic of the underlying algebra. For \ar the marks are $n_i = 1$ for all $i = 1,\ldots,r$. The parameter $m$ sets a mass scale, while $\beta$ is the coupling constant. When considering solitons, as is the case here, the coupling constant $\beta$ is usually taken to be imaginary \cite{Holl91,OTU92}.
\\
\\
Bowcock, Corrigan and Zambon, having first found a Lagrangian for a sine-Gordon defect \cite{BCZ03}, were able to generalise and find the type I defect for \ar \cite{BCZ04}
\begin{align}
\mathcal{L} = \theta (-x) \mathcal{L}_u + \theta (x) \mathcal{L}_v + \delta (x)\left( \frac{1}{2}u A \dot{u} + u B \dot{v} + \frac{1}{2}v A \dot{v} - D(u,v) \right) \; . \label{defect}
\end{align}
\\
Equation \eqref{defect} describes a defect located at $x=0$, where $B$ and $A = 1-B$ are constant matrices, while $\mathcal{L}_u$ and $\mathcal{L}_v$ are bulk \ar Lagrangians of the form \eqref{bulk} for the fields $u$ and $v$ respectively. The defect potential is given by
\begin{align}
D(u,v) = \frac{m}{\beta^2} e^{-\eta} \sum_{j=0}^r e^{\frac{1}{2} \beta \alpha_j \cdot \left( B^T u + B v \right)} + \frac{m}{\beta^2} e^{\eta} \sum_{j=0}^r e^{\frac{1}{2} \beta \alpha_j \cdot B (u-v)} \label{dpot}
\end{align}
where the parameter $\eta$ is identified as the `rapidity' of the defect. The parameter $\eta$ does transform as a rapidity under Lorentz boosts \cite{BCZ05}, however, a non-zero rapidity does not preclude the defect from being stationary. It is clear then, that along with $\eta$, the defect is fully specified by the constant matrix $B$, which has two relevant solutions for $r \geq 2$
\begin{align}
B_1 = 2 \sum_{j=1}^r \left( \lambda_j - \lambda_{j+1}\right) \lambda_j^T \label{bone}
\end{align}
and
\begin{align}
B_r = 2 \sum_{j=1}^r \left( \lambda_j - \lambda_{j-1}\right) \lambda_j^T \label{btwo}
\end{align}
where $\{ \lambda_i \}$ are the fundamental highest weights of \ar which satisfy $\lambda_i \cdot \alpha_j = \delta_{ij}$ for $i,j = 1,\ldots,r$ and $\lambda_0 = 0$. The identification made in this paper is that \eqref{bone} specifies a \emph{species 1} defect while \eqref{btwo} specifies a \emph{species $r$} defect. The type II defects of Corrigan and Zambon \cite{CZ10b} are then considered to be composite defects, consisting of a species 1 and a species $r$ defect combined\footnote{Previous literature \cite{CZ09b,CZ10b,Rob} refers to this as `fusing' defects. This nomenclature is avoided in this paper as fusing here refers to the fusing rules.} and given the same rapidity - this explains the foldability of such defects \cite{Rob} as it is analogous to how the solitons of the folded theory arise.
\\
\\
The classical solitons of \ar can be described in terms of Hirota tau functions \cite{Holl91} and the effect of left-to-right transmission through a defect was considered in \cite{BCZ04,CZ07}. For a species $p$ soliton\footnote{The possible species are $p = 1,\ldots,r$ which encompass what are elsewhere referred to as the `soliton' representations and the `anti-soliton' representations (e.g. \cite{CZ09a}). The anti-soliton of a species $p$ soliton is a species $h-p$ soliton where $h=r+1$ is the Coxeter number of the algebra.} passing through a species 1 defect the tau functions of $v$ pick up a delay factor of
\begin{align}
z^p_1 (\theta - \eta) = \frac{i e^{\eta - \theta} + \omega^{\frac{p}{2}}}{i e^{\eta - \theta} + \omega^{-\frac{p}{2}}} \label{del}
\end{align}
where $\theta$ is the rapidity of the soliton and $\omega = e^{\frac{2 \pi i}{h}}$, $h = r+1$ being the Coxeter number of \arf. 
\\
\\
The quantum solitons can be viewed as operators in the Faddeev--Zamolodchikov algebra \cite{Gand}. A species $p$ soliton with topological charge labelled by $i$ and rapidity $\theta$ has the operator denoted by ${}^p A_i (\theta)$. Defects can viewed in a similar manner \cite{CZ07}, so a species $q$ defect with topological charge $\alpha$ and rapidity $\eta$ is denoted by ${}_q D_{\alpha} (\eta)$. Thus, the transmission process has the algebra
\begin{align}
{}^p A_i (\theta) {}_q D_{\alpha}(\eta) = {}^p_q T_{i \alpha}^{n \lambda}(\theta - \eta) {}_q D_{\lambda}(\eta) {}^p A_n (\theta) \label{fzdef}
\end{align}
and the transmission is specified by the $T$-matrix ${}^p_q T_{i \alpha}^{n \lambda}(\theta - \eta)$. Note that, as expected \cite{DMS94b}, reflection is absent from this process.
\\
\\
Section \ref{classfuse} explains the classical fusing process for solitons before treating the defects analogously. The key quantity is the delay factor picked up by the soliton being transmitted through the defect. Section \ref{quantfuse} outlines the quantum fusing rules for solitons and postulates the analogous fusing rules for defects. Section \ref{newresult} uses the quantum fusing rules to help find a new transmission matrix in \athf. The conclusions and outlook can be found in section \ref{disgust}. Calculations for finding the transmission matrix of section \ref{newresult} can be found in the appendices.

\section{Classical fusing rules} \label{classfuse}
In this section a description of the classical solitons of \ar and their fusing rules is given. This is applied to the transmission of solitons through defects giving the delay factors - these delay factors conveniently illustrate both soliton and defect fusing rules.

\subsection{Classical solitons and transmission}

Using Hirota tau functions \cite{Hiro} a classical description of the \ar solitons can be found \cite{Holl91}. For \ar soliton solutions the field takes the form
\begin{align}
u &= -\frac{1}{\beta}\sum_{j=1}^r \alpha_j \ln \left( \frac{\tau_j}{\tau_0} \label{solanz} \right)
\end{align}
with $\{ \tau_j \}$ the Hirota tau functions. There are $r$ species of fundamental soliton in \ar, with the species $p$ single soliton given by
\begin{align}
\tau_j &= 1 + \omega^{pj} E_p \label{onesol}
\end{align}
where $\omega = e^{\frac{2 \pi i}{h}}$, as it is in \eqref{del}, and the spacetime dependence of the soliton is encapsulated in
\begin{align}
E_p &= e^{a_p x - b_p t + c_p} \; . \label{ep}
\end{align}
\\
In \eqref{ep}, $a_p = m_p \cosh \theta$ and $b_p = m_p \sinh \theta$ where $m_p = 2 m \sin \left( \frac{\pi p}{h} \right)$ and $\theta$ is the rapidity of the soliton. The mass of the soliton is given by $M_p = \frac{2 h}{|\beta^2|} m_p$. The real part of $c_p$ relates to the centre of mass of the soliton while the imaginary part relates to the topological charge. The possible topological charges of the soliton lie within the weight space of the $p$-th fundamental representation of $a_r$, but in many cases not all of the weights correspond to soliton charges at the classical level \cite{McGh}. \\
\\
A two soliton solution, say a species $p_1$ soliton and a species $p_2$ soliton, is given by the tau functions \cite{Holl91}
\begin{align}
\tau_j = 1 + \omega^{p_1 j} E_{p_1} + \omega^{p_2 j} E_{p_2} + A^{(p_1 p_2)} \omega^{(p_1 + p_2)j} E_{p_1} E_{p_2} \label{twosol}
\end{align}
which has an interaction parameter
\begin{align}
A^{(p_1 p_2)} = - \frac{ (a_{p_1} - a_{p_2})^2 - (b_{p_1} - b_{p_2})^2 - m_{p_1 - p_2} }{(a_{p_1} + a_{p_2})^2 - (b_{p_1} + b_{p_2})^2 - m_{p_1 + p_2}} \; . \label{intparam}
\end{align}
\\
The tau functions \eqref{twosol} can, in the correct circumstances, describe a single soliton of species $p_3 = p_1 + p_2 \, (\text{mod }h)$ \footnote{If $p_3 = 0$ then it is the trivial solution.}, which occurs when the constituent solitons are placed at the same location with a rapidity difference of $\theta_1 - \theta_2 = \pm i \frac{\pi (p_1 + p_2)}{h}$, i.e., $i$ times the fusing angle \cite{OTU93}. Under such circumstances the interaction parameter \eqref{intparam} has a pole, so by shifting $E \to E A^{-\frac{1}{2}}$ it can be seen that \eqref{twosol} reduces to \eqref{onesol} with species $p_3$ \cite{Hall}. Note that the fusing process described here can violate topological charge conservation \cite{McGh}.
\\
\\
Consider now the species 1 defect with Lagrangian given by \eqref{defect} and $B$ given by \eqref{bone}. The Euler--Langrange equations at the defect ($x=0$) give
\begin{align}
u' &= A \dot{u} + B \dot{v} - \nabla_u D \nonumber \\
v' &= -A \dot{v} + B^T \dot{u} + \nabla_v D \label{eles} \; .
\end{align}
\\
Using \eqref{eles} it can be shown that a single soliton of species $p$ moving through the defect is transmitted but picks up the delay factor \eqref{del}, so
\begin{align}
\tau_j (v) = 1 + \omega^{pj} z_1^p(\theta - \eta) E_p \label{taudel} \; .
\end{align}
\\
Now consider the case where the soliton is a species 2 soliton, which can also be viewed as a two-soliton solution involving two species 1 solitons at the same location with rapidities $\theta_1 = \theta + \frac{i \pi}{h}$ and $\theta_2 = \theta - \frac{i \pi}{h}$. Since the individual solitons are delayed independently by the defect it must be the case that
\begin{align}
z_1^2 (\theta - \eta) = z_1^1 \left( (\theta + \tfrac{i \pi}{h}) - \eta \right) z_1^1 \left( (\theta - \tfrac{i \pi}{h}) - \eta \right) \label{fusedel}
\end{align}
so the soliton fusing rules are clearly seen in the delay factors.

\subsection{Defect fusing rules}

The fundamental solitons of the $a_r^{\! (1)}$ ATFT can be thought of as carrying the fundamental representations of the $a_r$ algebra, since the topological charge of a species $p$ soliton is a weight of the $p$-th fundamental representation of the algebra. The topological charge of an $a_r^{\! (1)}$ defect can be anywhere in the root space of $a_r$, suggesting that defects have an association with certain infinite-dimensional representations of $a_r$. The global symmetry of \ar should therefore determine the fusing angles and representations of the defects. The fusing angles are taken here to be the same as those of the solitons. Evidence supporting this identification comes from the closure of the defect bootstrap in terms of the soliton delay factors in equation \eqref{fusegen}. In particular a species 2 defect should arise when species 1 defects are combined with a rapidity difference of $i$ times the fusing angle, $\eta_1 - \eta_2 = \pm i \frac{2 \pi}{h}$.
\\
\\
In a vacuum configuration, with $u$ and $v$ in the same representation, the species 1 defect (and the species $r$ defect) possesses an energy and a momentum given by $(E,P) = (\frac{2hm}{\beta^2}\cosh \eta , -\frac{2hm}{\beta^2}\sinh \eta)$ suggesting a mass of
\begin{align}
\mathcal{M}_1 &= \frac{2 h m}{|\beta^2|} \; . \label{defmass}
\end{align}
\\
Thus, taking $\eta_1 = \eta - \frac{i \pi}{h}$ and $\eta_2 = \eta + \frac{i \pi}{h}$ the species 2 soliton will have a mass of
\begin{align*}
\mathcal{M}_2 = \frac{4 h m}{\beta^2}\cos \left( \frac{\pi}{h} \right) = 2\cos \left( \frac{\pi}{h} \right) \mathcal{M}_1
\end{align*}
so it is notable that the mass ratios of the defects are the same as those of the solitons and of the elementary excitations: $\frac{\mathcal{M}_1}{\mathcal{M}_2} = \frac{M_1}{M_2} = \frac{m_1}{m_2}$ \cite{BCDS,OTU92}.
\\
\\
It is clear that a single soliton is delayed independently by different defects so the delay factor through a species 2 defect must be
\begin{align}
z_2^1 (\theta - \eta) = z_1^1 \left( \theta - (\eta - \tfrac{i \pi}{h}) \right) z_1^1 \left( \theta - (\eta + \tfrac{i \pi}{h}) \right) \label{fusedef}
\end{align}
and so $z_2^1$ clearly matches $z_1^2$ as given in \eqref{fusedel}. In general a species $q_1$ and a species $q_2$ defect can fuse to form a species $q_3 \, (\text{mod }h)$ defect with delay factor
\begin{align}
z_{q_3}^p (\theta - \eta) = z_{q_1}^p \left( \theta - \eta + \tfrac{i \pi q_2}{h} \right) z_{q_2}^p \left( \theta - \eta - \tfrac{i \pi q_1}{h} \right) \; . \label{fusegen}
\end{align}
\\
The idea of an anti-defect can be introduced in analogy to how antisolitons are related to solitons. An anti-defect and a defect give the reciprocal delay factors and should annihilate when combined. It is evident from the fusing rules that the anti-defect of a species $q$ defect with rapidity $\eta$ is a species $h-q$ defect with rapidity $\eta \pm i \pi$.
\\
\\
Note also that:
\begin{itemize}

\item The Lagrangian of a species 2 defect is expected to arise by taking the Lagrangian for two species 1 defects and combining the defects. The species 2 defect Lagrangian is then generally of type II form, in that it possesses an auxiliary field. The exception to this is in $a_2^{\! (1)}$ where the species 2 defect is known to have a type I Lagrangian - this can be recovered from the type II description with a particular identification of the auxiliary field.

\item In \ath the species 2 defect gives the same delay factor to the species 1 soliton as it does to the species 3 soliton. This is analogous to how the species 2 soliton delays the other solitons.

\end{itemize}

\section{Quantum fusing rules} \label{quantfuse}

In this section a description is given of the \ar solitons in the quantum context along with their fusing rules. Defect fusing rules are then described in an analogous way.

\subsection{Quantum solitons and transmission}

The scattering of solitons in \ar can be conveniently framed in terms of the Faddeev--Zamolodchikov algebra \cite{ZZ,Gand}. There are $r$ possible species of single soliton, with a species $p$ soliton possessing as its topological charge one of the weights of the $p$-th fundamental representation of $a_r$. In this section it is assumed that all of the weights of the representation appear as topological charges in the quantum theory. That this does not match up with the classical theory, where not all weights are charges, has long been known \cite{Holl92} and remains an unresolved issue.
\\
\\
One can represent a species $p$ soliton possessing rapidity $\theta$ and topological charge label $i$ as an operator ${}^{p  \! \!}A_i (\theta)$; then the algebra describing the scattering of two solitons is
\begin{align}
{}^{p_1  \! \!}A_j (\theta_1) \, {}^{p_2  \! \!}A_k (\theta_2) = {}^{p_1 p_2} S_{jk}^{mn} (\theta_1 - \theta_2) \, {}^{p_2  \! \!}A_m (\theta_2) \, {}^{p_1 \! \!}A_n (\theta_1) \; . \label{FZ}
\end{align} 
\\
In \eqref{FZ} the in-state, ${}^{p_1  \! \!}A_j (\theta_1) \, {}^{p_2  \! \!}A_k (\theta_2)$, is related to the out-state, ${}^{p_2  \! \!}A_m (\theta_2) \, {}^{p_1 \! \!}A_n (\theta_1)$, by the $S$-matrix. The $S$-matrices for the \ar solitons were originally postulated by Hollowood \cite{Holl92}. 
\\
\\
The fusing rules for the solitons can be seen in this algebra. In particular, the operator for a species 2 soliton may be written in terms of species 1 operators \cite{CZ07}
\begin{align}
{}^{2 \! \!} A_{(jk)} (\theta) = c^{(jk)} \, {}^{1 \! \!} A_j (\theta - \tfrac{i \pi}{h}) {}^{1 \! \!} A_k (\theta + \tfrac{i \pi}{h}) + (j \leftrightarrow k) \label{optwo}
\end{align}
where $c^{(jk)}$ are the fusing couplings of the theory for this process. Similarly a species $p_1$ and a species $p_2$ soliton can combine to form a species $p_3 = p_1 + p_2 \, (\text{mod }h)$ soliton
\begin{align}
{}^{p_3 \! \!} A_{(jk)} (\theta) = c^{(jk)}(p_1,p_2) \, {}^{p_1 \! \!} A_j (\theta - \tfrac{i \pi p_2}{h}) {}^{p_2 \! \!} A_k (\theta + \tfrac{i \pi p_1}{h}) + (j \leftrightarrow k) \label{opp}
\end{align}
where the couplings involved $\{ c^{(jk)} \}$ depend on the species of solitons being fused. \\
\\
By making use of \eqref{opp} and \eqref{FZ} the different scattering matrices can be found in terms of the $S$-matrices for species 1 solitons. 
\\
\\
Defects may also be factored into the Faddeev--Zamolodchikov algebra by introducing a defect operator \cite{CZ07}. A defect of species $q$ and rapidity $\eta$ carrying a topological charge of $\alpha$ has the operator ${}_q D_{\alpha}(\eta)$, so the left-to-right transmission of a species $p$ soliton (Re$(\theta) > 0$) through such a defect has the algebra
\begin{align}
{}^{p \! \!}A_i (\theta) \, {}_q D_{\alpha}(\eta) = {}^p_q T_{i \alpha}^{n \lambda}(\theta - \eta) {}_q D_{\lambda}(\eta) \, {}^{p \! \!}A_n (\theta) \; . \label{FZT}
\end{align}
\\
In \eqref{FZT} the in-state, ${}^{p \! \!}A_i (\theta) \, {}_q D_{\alpha}(\eta)$ which is where the soliton is to the left of the defect, is related to the out state, $D_{\lambda}(\eta) \, {}^{p \! \!}A_n (\theta)$, by means of the $T$-matrix. The defects considered in this paper are considered to be in their `ground state', meaning that they are stable and do not change the species of the transmitted soliton; as such, the possible topological charges of the defects, for every species of defect, lie in the root lattice of the \ar theory under consideration. Only for species 1 and species $r$ defects have any of these $T$-matrices appeared in the literature \cite{CZ07,CZ09a}. \\
\\
Using \eqref{optwo} one can find the transmission matrix for a species 2 soliton from that of a species 1 soliton in a simple manner
\begin{align}
{}^{2}T_{(jk)}^{(ab)}(\theta - \eta) c^{(ab)} = c^{(jk)} {}^1 T_j^a (\theta - \eta - \tfrac{i \pi}{h}) {}^1 T_k^b (\theta - \eta + \tfrac{i \pi}{h}) + (j \leftrightarrow k) \label{fuset}
\end{align}
with no sum implied.

\subsection{Quantum defect fusing rules} \label{fusey}

It is proposed now that the operator for a species 2 defect can similarly be written in terms of the operators for species 1 defects, i.e.,
\begin{align}
{}_2 D_{\alpha}(\eta) = \sum_{\beta, \gamma, \beta + \gamma = \alpha} d_{11}^{\beta, \gamma} {}_1 D_{\beta}(\eta + \tfrac{i \pi}{h})  \, {}_1 D_{\gamma}(\eta - \tfrac{i \pi}{h}) \label{doptwo}
\end{align}
where $\{ d^{\beta, \gamma} \}$ are the defect fusing couplings. As in section \ref{classfuse}, the fusing angles for defects have been taken to be the same as the fusing angles in the analogous soliton fusing process. By using the defect fusing equation \eqref{doptwo} with \eqref{FZT} the transmission matrix for a soliton through a species 2 defect can be written in terms of the transmission matrices through species 1 defects as
\begin{align}
{}_2 T_{i \alpha}^{n \lambda} (\theta - \eta) d_{11}^{\delta, \epsilon} = \sum_{\beta, \gamma, j} d_{11}^{\beta, \gamma} {}_1 T_{i \beta}^{j \delta}(\theta - \eta - \tfrac{i \pi}{h}) {}_1 T_{i \gamma}^{j \epsilon}(\theta - \eta + \tfrac{i \pi}{h}) \label{proanz}
\end{align}
where $\beta + \gamma = \alpha$ and $\delta + \epsilon = \lambda$. This equation could be significant in the discovery of new transmission matrices, but has limited use in generating new solutions until the ratios of the defect fusing couplings are known. However, since the transmission matrices for solitons through species 1 defects are known \cite{CZ09a}, equation \eqref{proanz} can at least provide an ansatz for the form of the transmission matrix for the species 2 defect - this is considered for the simplest new case, \athf, in the next section. \\
\\
In general it is expected that a species $q_1$ and a species $q_2$ defect should be able to fuse to form a species $q_3 = q_1 + q_2 \, (\text{mod }h)$ defect\footnote{If $q_3 = 0$ then this is where a defect and anti-defect have annihilated so there is no defect there, although this is not obvious from the Lagrangian.}, so
\begin{align}
{}_{q_3}T_{i \alpha}^{n \lambda} d_{q_1 q_2}^{\delta, \epsilon} = \sum_{\beta, \gamma, j} d_{q_1 q_2}^{\beta, \gamma} \; {}_{q_1}T_{i \beta}^{j \delta}(\theta - \eta - \tfrac{i \pi q_2}{h}) \,  {}_{q_2}T_{j \gamma}^{n \epsilon}(\theta - \eta + \tfrac{i \pi q_1}{h}) \label{genfuse}
\end{align}
where again $\beta + \gamma = \alpha$ and $\delta + \epsilon = \lambda$. If the ratios of the defect fusing couplings were known in general it would be possible to write all fundamental defect transmission matrices in terms of species 1 transmission matrices.
\\
\\
The defect fusing couplings $\{ d^{\beta , \gamma} \}$ should be determinable from the quantum group symmetry of the system. Since there are an infinite number of choices for the charges $\beta$ and $\gamma$ it would appear that the representations of interest are infinite-dimensional, but it is not clear which infinite-dimensional representations should be considered, hence the quantum group approach is not considered here.

\section{A new defect in \ath} \label{newresult}

In this section the defect fusing idea is applied to the case of \athf, which is the ATFT with the lowest rank for which the fusing rules give a previously unconsidered defect. A transmission matrix is found for the species 2 defect.

\subsection{Transmission matrix ansatz}

As noted in section \ref{fusey}, the defect fusing equation \eqref{proanz}, which gives the species 2 defect transmission matrix in terms of species 1 defect transmission matrices, appears to have limited use while the couplings $\{ d^{\beta, \gamma} \}$ remain unknown. However, with the assumption that the couplings depend only on topological charge and not on rapidity, \eqref{proanz} can be used to get an ansatz for the species 2 defect $T$-matrix. \\
\\
The `ground state' species 1 defect transmission matrices in \ath (and \ar generally) are known; with the species 1 soliton they are \cite{CZ09a}
\begin{align}
{}^1_1 T_{i \alpha}^{i \lambda}(\theta - \eta) = g^1 (\theta - \eta) Q^{\lambda \cdot l_1} \delta_{\alpha}^{\lambda} & & {}^1_1 T_{i \alpha}^{(i-1) \lambda}(\theta - \eta) = g^1 (\theta - \eta) \hat{x} \delta_{\alpha}^{\lambda + \alpha_{i-1}} & & i = 1,2,3,4 \label{toneone}
\end{align}
where the situation $i-1 = 0$ should be taken as $i-1 = 4$. The weights of the first representation are given by $l_i = \frac{3}{4}\alpha_i + \frac{1}{2}\alpha_{i+1} + \frac{1}{4}\alpha_{i+2}$ where the labels on the roots are modulo $h=4$; $Q = - e^{i \pi \gamma}$ with $\gamma = \frac{4 \pi}{\beta^2} - 1$ where $\beta$ is the bulk coupling appearing in \eqref{pot} and \eqref{dpot}. The quantity $\hat{x} = e^{\gamma (\theta - \eta - \frac{i \pi}{2})}$ relates to the likelihood of the soliton exchanging topological charge with the defect. The transmission matrix has a prefactor given by \cite{CZ09a}
\begin{align*}
g^1 (\theta - \eta) = \frac{\hat{x}^{-\frac{1}{2}}}{2 \pi} \Gamma (\tfrac{1}{2} + \tfrac{3}{2}\gamma - z) \prod_{k=1}^{\infty} \frac{ \Gamma (\frac{1}{2} + (4k + \frac{3}{2})\gamma - z) \Gamma (\frac{1}{2} + (4k - \frac{5}{2})\gamma + z) }{ \Gamma (\frac{1}{2} + (4k - \frac{3}{2})\gamma - z) \Gamma (\frac{1}{2} + (4k - \frac{3}{2})\gamma + z) \vphantom{\frac{T}{T}} }
\end{align*}
where $z = \frac{2i\gamma \left(\theta - \eta - \frac{i \pi}{2} \right)}{\pi}$. \\
\\
Using this result for ${}^1_1 T$ in \eqref{proanz} the ansatz for the transmission of a species 1 soliton through a species 2 defect, ${}^1_2 T$, obtained is
\begin{align}
& {}^{1}_{2}T_{\alpha}^{\lambda} (\theta - \eta) = g^2 (\theta - \eta) \nonumber \\
& \; \; \times  \left(
\begin{array}{cccc}
Q^{\lambda \cdot l_1}\delta_{\alpha}^{\lambda} & 0 & \hat{x}^2 b_{13}(\lambda)\delta_{\alpha}^{\lambda -  \alpha_1 - \alpha_2} & \hat{x} a_{14}(\lambda) \delta_{\alpha}^{\lambda + \alpha_0} \\
\hat{x} a_{21}(\lambda) \delta_{\alpha}^{\lambda + \alpha_1} & Q^{\lambda \cdot l_2}\delta_{\alpha}^{\lambda} & 0 & \hat{x}^2 b_{24}(\lambda)\delta_{\alpha}^{\lambda -  \alpha_2 - \alpha_3} \\
\hat{x}^2 b_{31}(\lambda)\delta_{\alpha}^{\lambda +  \alpha_1 + \alpha_2} & \hat{x} a_{32}(\lambda) \delta_{\alpha}^{\lambda + \alpha_2} & Q^{\lambda \cdot l_3}\delta_{\alpha}^{\lambda} & 0  \\
0 & \hat{x}^2 b_{42}(\lambda)\delta_{\alpha}^{\lambda +  \alpha_2 + \alpha_3} & \hat{x} a_{43}(\lambda) \delta_{\alpha}^{\lambda + \alpha_3} & Q^{\lambda \cdot l_4}\delta_{\alpha}^{\lambda}
\end{array}
\right)  \label{ttwoone}
\end{align}
where the prefactor is $g^2 (\theta - \eta) = g^1 (\theta - \eta - \tfrac{i \pi}{4}) g^1 (\theta - \eta + \tfrac{i \pi}{4})$, while the functions $\{ a_{ij}(\lambda) \}$ and $\{ b_{ij}(\lambda) \}$ are unknown but depend only on the topological charges of the defect and soliton and not on the rapidities.
\\
\\
Note that \eqref{ttwoone} need only be given in terms of the species 1 soliton as, once a consistent solution has been found for ${}^1_2 T$, the transmission matrices for the other solitons must follow from the soliton fusing rules. In fact, soliton fusing can be used to constrain the form of $a_{ij}(\lambda)$ and $b_{ij}(\lambda)$ since the topological charge of the species 2 soliton can be formed in two ways, but the constraints obtained are just a subset of those arising from the triangle relations \eqref{triangle}.

\subsection{Constraining the $T$-matrix}

Two types of constraint are used in this paper to determine the $T$-matrix \eqref{ttwoone}. The first is a form of Yang--Baxter equation for two solitons and a defect, known as the triangle relations
\begin{align}
{}^{11}S_{jk}^{mn}(\theta_1 - \theta_2) {}_2^1 T_{n \alpha}^{t \beta}(\theta_1 - \eta) {}_2^1 T_{m \beta}^{s \lambda}(\theta_2 - \eta) = {}_2^1 T_{k \alpha}^{m \beta}(\theta_2 - \eta) {}_2^1 T_{j \beta}^{n \lambda}(\theta_1 - \eta) {}^{11}S_{nm}^{st}(\theta_1 - \theta_2) \label{triangle}
\end{align}
where $m$, $n$ and $\beta$ are summed over. the triangle relations were previously used by Corrigan and Zambon to find $T$-matrices for other defects \cite{CZ07,CZ09a,CZ10b} though quantum group methods can also be used \cite{CZ10b}. The triangle relations do not constrain the prefactor $g^2(\theta - \eta)$ but this is not an issue here as it's already fully determined by the fusing rules and previous results \cite{CZ09a}.
\\
\\
The second method to constrain the $T$-matrix uses the crossing and unitarity conditions. The unitarity condition is \cite{CZ07,CZ09a}
\begin{align*}
{}^1_2 T_{i \alpha}^{j \beta} (\theta - \eta) {}^1_2 \tilde{T}_{j \beta}^{n \lambda} (\eta - \theta) = \delta_i^n \delta_{\alpha}^{\lambda}
\end{align*}
where $\tilde{T}$ is the transmission matrix for a soliton moving right-to-left through the defect. In analogy to the species 2 soliton, it is expected that the species 2 defect of \ath is self-conjugate, in that its anti-defect is another species 2 defect with opposite topological charge, so there will be a crossing relation
\begin{align*}
{}_2^1 T_{i \alpha}^{n \lambda}(\theta - \eta) = {}_2^1 \tilde{T}_{i (-\lambda)}^{n (-\alpha)}(i\pi + \eta - \theta) \; .
\end{align*}
\\
These crossing and unitarity relations combine to give
\begin{align}
{}_2^1 T_{i \alpha}^{j \beta}(\theta - \eta) {}_2^1 T_{j (-\lambda)}^{n (-\beta)}(\theta - \eta + i\pi) = \delta_i^n \delta_{\alpha}^{\lambda} \; . \label{cruni}
\end{align}
\\
Note that \eqref{cruni} can be interpreted as a description of the transmission of a soliton through a (species 2) defect immediately followed by transmission through an anti-defect, the combined effect being trivial.
\\
\\
The triangle relations \eqref{triangle} and crossing-unitarity relations \eqref{cruni} are used in appendix \ref{appa} to constrain \eqref{ttwoone}.

\subsection{Solutions}

Two solutions are found in appendix \ref{appa} which satisfy \eqref{triangle} and \eqref{cruni}. These, written in their most symmetric form, are
\begin{align}
b_{ij}(\lambda) &= Q^{-\frac{1}{2} \lambda \cdot (l_i + l_j)} \nonumber \\
a_{ij}(\lambda) &= Q^{\frac{1}{2} \lambda \cdot (l_i + l_j)} \left( Q^{-\frac{1}{4} + \frac{1}{2}\lambda \cdot (l_i + l_{i+2})} + Q^{\frac{1}{4} - \frac{1}{2}\lambda \cdot (l_i + l_{i+2})} + (-1)^{i+1}\sqrt{2} \right) \label{sol1}
\end{align}
and
\begin{align}
b_{ij}(\lambda) &= Q^{-\frac{1}{2} \lambda \cdot (l_i + l_j)} \nonumber \\
a_{ij}(\lambda) &= Q^{\frac{1}{2} \lambda \cdot (l_i + l_j)} \left( Q^{-\frac{1}{4} + \frac{1}{2}\lambda \cdot (l_i + l_{i+2})} + Q^{\frac{1}{4} - \frac{1}{2}\lambda \cdot (l_i + l_{i+2})} + (-1)^{i}\sqrt{2} \right) \; . \label{sol2}
\end{align}
\\
Rescaling and unitary transformations are considered in appendix \ref{appb} but no inequivalent solutions are found. There is no indication here that one of \eqref{sol1} and \eqref{sol2} should be favoured as the solution, further analysis will be required to determine whether or not both solutions should be considered as valid.

\section{Discussion} \label{disgust}

\subsection{Conclusions}

This paper has made the case for the existence of defect fusing rules in the \ar ATFTs. When defects are approached as being a kind of particle the possibility of defect fusing rules is an issue which must be addressed. Reasons to view defects as particles include: the existence of anti-defects, energy-momentum being associated to defects and the observation that certain defects can mimic solitons.
\\
\\
The main result of the paper is the application of the defect fusing rule idea at the quantum level to find a new transmission matrix in \ath. The fusing rule idea naturally identifies this as a transmission matrix for the hitherto unconsidered species 2 defect.

\subsection{Outlook}

The existence of defect fusing rules allows for a more systematic study of affine Toda defects in the future. A step in systematising the process is the examination of \eqref{proanz} in the $a_2^{\! (1)}$ case, where both the species 1 and species 2 transmission matrices are known. In that case \eqref{proanz} reduces to a set of equations to determine the ratios of the defect fusing couplings. \\
\\
In the absence of the coupling ratios new transmission matrices can be found on a case-by-case basis. The most similar case to the species 2 defect of \ath found in this paper is the species 3 defect of $a_5^{\! (1)}$, which is also expected to be self-conjugate. Similar conditions for the transmission matrices for this defect can be found to those in appendix \ref{appa} but they are significantly more complicated to solve.
\\
\\
Whilst no integrable defects have been found for simply laced ATFTs other than the \ar series, defect fusing rules mean that progress can be made once the `basic' defects are found. Coupled to the idea of folding defect configurations \cite{Rob}, defect fusing rules should allow contact with all defects in all of the ATFTs.
\\
\\
Defect fusing rules could also play a part in the study of defect-defect scattering. Provided a scattering matrix can be found for the interaction of two species 1 defects in \ar, the fusing rules could be used to generate the scattering matrices of other defect species.

\subsection*{Acknowledgements}

The author wishes to thank Peter Bowcock and Ed Corrigan for their comments and suggestions. This work was supported by an STFC studentship.

\appendix

\section{Solving the $T$-matrix constraints}

In this section the $T$-matrix constraints \eqref{triangle} and \eqref{cruni} are used on the species 2 defect ansatz \eqref{ttwoone} to get the solutions \eqref{sol1} and \eqref{sol2}.
\\
\\
The first set of constraints is the triangle relations \eqref{triangle}
\begin{align*}
{}^{11}S_{jk}^{mn}(\theta_1 - \theta_2) {}_2^1 T_{n \alpha}^{t \beta}(\theta_1 - \eta) {}_2^1 T_{m \beta}^{s \lambda}(\theta_2 - \eta) = {}_2^1 T_{k \alpha}^{m \beta}(\theta_2 - \eta) {}_2^1 T_{j \beta}^{n \gamma}(\theta_1 - \eta) {}^{11}S_{nm}^{st}(\theta_1 - \theta_2)
\end{align*}
where $m$ and $n$ are summed over on both sides\footnote{The intermediate defect topological charge $\beta$ also takes multiple values but their relation to the outgoing charge $\lambda$ is fixed by the values of $m$ and $n$.}. As well as the ansatz \eqref{ttwoone} for ${}_2^1 T$, the $S$-matrix between two species 1 solitons is used. It is given by \cite{CZ07}
\begin{align*}
{}^{11}S_{jj}^{jj}(\theta_1 - \theta_2) &= \rho (\theta_1 - \theta_2 ) \left( Q^{-1} \frac{\hat{x}_1^2}{\hat{x}_2^2} - Q \frac{\hat{x}_2^2}{\hat{x}_1^2} \right) \\
{}^{11}S_{jk}^{kj}(\theta_1 - \theta_2) &= \rho (\theta_1 - \theta_2 ) \left( \frac{\hat{x}_1^2}{\hat{x}_2^2} - \frac{\hat{x}_2^2}{\hat{x}_1^2} \right) \quad j \neq k \\
{}^{11}S_{jk}^{jk}(\theta_1 - \theta_2 ) &= \rho (\theta_1 - \theta_2 ) \left( Q^{-1} - Q \right) 
\begin{cases}
 \left(\dfrac{\hat{x}_1}{\hat{x}_2}\right)^{(2 - |l|)} |_{l = j - k <0} \\
 \left(\dfrac{\hat{x}_2}{\hat{x}_1}\right)^{(2 - |l|)} |_{l = j - k >0}
\end{cases}
\end{align*}
where $\frac{\hat{x}_1}{\hat{x}_2} = e^{\gamma (\theta_1 - \theta_2)}$ and again $Q = - e^{i \pi \gamma}$ with $\gamma = \frac{4 \pi}{\beta^2} - 1$. The prefactor $\rho (\theta_1 - \theta_2)$ is immaterial to this discussion as it appears as a common factor on both sides of \eqref{triangle} - an expression for $\rho (\theta)$ can be found in \cite{CZ07}. The triangle relations in this case are a set of $4^4 = 256$ conditions but most of these are trivially satisfied, there are 28 non-trivial conditions which are:
\begin{align*}
a_{14}(\lambda) a_{21}(\lambda + \alpha_0) - a_{14}(\lambda + \alpha_1) a_{21}(\lambda)  &= \left(Q^{-1} - Q \right) b_{24}(\lambda)  Q^{\lambda \cdot l_1} \\
a_{21}(\lambda) a_{32}(\lambda + \alpha_1) - a_{21}(\lambda + \alpha_2) a_{32}(\lambda)  &= \left(Q^{-1} - Q \right) b_{31}(\lambda)  Q^{\lambda \cdot l_2} \\
a_{32}(\lambda) a_{43}(\lambda + \alpha_2) - a_{32}(\lambda + \alpha_3) a_{43}(\lambda)  &= \left(Q^{-1} - Q \right) b_{42}(\lambda)  Q^{\lambda \cdot l_3} \\
a_{43}(\lambda) a_{14}(\lambda + \alpha_3) - a_{43}(\lambda + \alpha_0) a_{14}(\lambda)  &= \left(Q^{-1} - Q \right) b_{13}(\lambda)  Q^{\lambda \cdot l_4} \tag{$A$} \label{cona}
\end{align*}
\begin{align*}
a_{14}(\lambda) b_{31}(\lambda + \alpha_0) &= a_{14}(\lambda + \alpha_1 + \alpha_2)b_{31}(\lambda) \\
a_{14}(\lambda) b_{42}(\lambda + \alpha_0) &= a_{14}(\lambda + \alpha_2 + \alpha_3)b_{42}(\lambda) \\
a_{21}(\lambda) b_{13}(\lambda + \alpha_1) &= a_{21}(\lambda - \alpha_1 - \alpha_2)b_{13}(\lambda) \\
a_{21}(\lambda) b_{42}(\lambda + \alpha_1) &= a_{21}(\lambda + \alpha_2 + \alpha_3)b_{42}(\lambda) \\
a_{32}(\lambda) b_{13}(\lambda + \alpha_2) &= a_{32}(\lambda - \alpha_1 - \alpha_2)b_{13}(\lambda) \\
a_{32}(\lambda) b_{24}(\lambda + \alpha_2) &= a_{32}(\lambda - \alpha_2 - \alpha_3)b_{24}(\lambda) \\
a_{43}(\lambda) b_{24}(\lambda + \alpha_3) &= a_{43}(\lambda - \alpha_2 - \alpha_3)b_{24}(\lambda) \\
a_{43}(\lambda) b_{31}(\lambda + \alpha_3) &= a_{43}(\lambda + \alpha_1 + \alpha_2)b_{31}(\lambda) \tag{$B$} \label{conb}
\end{align*}
\begin{align*}
a_{14}(\lambda) a_{32}(\lambda + \alpha_0) &= a_{14}(\lambda + \alpha_2) a_{32}(\lambda) \\
a_{21}(\lambda) a_{43}(\lambda + \alpha_1) &= a_{21}(\lambda + \alpha_3) a_{43}(\lambda) \tag{$C$} \label{conc}
\end{align*}
\begin{align*}
b_{13}(\lambda) b_{24}(\lambda - \alpha_1 - \alpha_2) &= b_{13}(\lambda - \alpha_2 - \alpha_3) b_{24}(\lambda) \\
b_{13}(\lambda) b_{42}(\lambda - \alpha_1 - \alpha_2) &= b_{13}(\lambda + \alpha_2 + \alpha_3) b_{42}(\lambda) \\
b_{31}(\lambda) b_{24}(\lambda + \alpha_1 + \alpha_2) &= b_{31}(\lambda - \alpha_2 - \alpha_3) b_{24}(\lambda) \\
b_{31}(\lambda) b_{42}(\lambda + \alpha_1 + \alpha_2) &= b_{31}(\lambda + \alpha_2 + \alpha_3) b_{42}(\lambda) \; .  \tag{$D$} \label{cond}
\end{align*}
\begin{align*}
b_{13}(\lambda) b_{31}(\lambda - \alpha_1 - \alpha_2) &= b_{13}(\lambda + \alpha_1 + \alpha_2) b_{31}(\lambda) \\
b_{24}(\lambda) b_{42}(\lambda - \alpha_2 - \alpha_3) &= b_{24}(\lambda + \alpha_2 + \alpha_3) b_{42}(\lambda) \tag{$E$} \label{cone}
\end{align*}
\begin{align*}
a_{14}(\lambda)b_{13}(\lambda + \alpha_0) &= Q a_{14}(\lambda - \alpha_1 - \alpha_2)b_{13}(\lambda) \\
a_{21}(\lambda)b_{24}(\lambda + \alpha_1) &= Q a_{21}(\lambda - \alpha_2 - \alpha_3)b_{24}(\lambda) \\
a_{32}(\lambda)b_{31}(\lambda + \alpha_2) &= Q a_{32}(\lambda + \alpha_1 + \alpha_2)b_{31}(\lambda) \\
a_{43}(\lambda)b_{42}(\lambda + \alpha_3) &= Q a_{43}(\lambda + \alpha_2 + \alpha_3)b_{42}(\lambda) \tag{$F$} \label{conf}
\end{align*}
\begin{align*}
a_{14}(\lambda)b_{24}(\lambda + \alpha_0) &= Q^{-1} a_{14}(\lambda - \alpha_2 - \alpha_3)b_{24}(\lambda) \\
a_{21}(\lambda)b_{31}(\lambda + \alpha_1) &= Q^{-1} a_{21}(\lambda + \alpha_1 + \alpha_2)b_{31}(\lambda) \\
a_{32}(\lambda)b_{42}(\lambda + \alpha_2) &= Q^{-1} a_{32}(\lambda + \alpha_2 + \alpha_3)b_{42}(\lambda) \\
a_{43}(\lambda)b_{13}(\lambda + \alpha_3) &= Q^{-1} a_{43}(\lambda - \alpha_1 - \alpha_2)b_{13}(\lambda) \tag{$G$} \; . \label{cong}
\end{align*}
\\
It is a difficult task to use the triangle relations alone to determine the solution to \eqref{ttwoone}; fortunately the self-conjugate property of the species 2 defect gives additional constraints via the crossing unitarity conditions \eqref{cruni}
\begin{align*}
{}_2^1 T_{i \alpha}^{j \beta}(\theta - \eta) {}_2^1 T_{j (-\lambda)}^{n (-\beta)}(\theta - \eta + i\pi) = \delta_i^n \delta_{\alpha}^{\lambda} \; . 
\end{align*}
\\
Note that the prefactor of \eqref{ttwoone} obeys
\begin{align*}
g^2 (\theta - \eta) g^2 (\theta - \eta + i \pi) = \frac{1}{1 + Q^2 \hat{x}^4}
\end{align*}
which is significant in the diagonal terms of \eqref{cruni}. The crossing-unitarity conditions are
\begin{align*}
b_{13}(\lambda) b_{31}(-\lambda) &= 1 \\
b_{24}(\lambda) b_{42}(-\lambda) &= 1   \tag{$A'$} \label{conap}
\end{align*}
\begin{align*}
a_{14}(\lambda) Q^{-\lambda \cdot l_1} &= a_{14}(-\lambda - \alpha_0 ) Q^{\lambda \cdot l_4} \\
a_{21}(\lambda) Q^{-\lambda \cdot l_2} &= a_{21}(-\lambda - \alpha_1 ) Q^{\lambda \cdot l_1} \\
a_{32}(\lambda) Q^{-\lambda \cdot l_3} &= a_{32}(-\lambda - \alpha_2 ) Q^{\lambda \cdot l_2} \\
a_{43}(\lambda) Q^{-\lambda \cdot l_4} &= a_{43}(-\lambda - \alpha_3 ) Q^{\lambda \cdot l_3} \tag{$B'$} \label{conbp}
\end{align*} 
\begin{align*}
a_{14}(\lambda) b_{31}(-\lambda + \alpha_3 ) &= Q^{-1} a_{32}(-\lambda + \alpha_3 ) b_{24}(\lambda) \\
a_{21}(\lambda) b_{42}(-\lambda + \alpha_0 ) &= Q^{-1} a_{43}(-\lambda + \alpha_0 ) b_{31}(\lambda) \\
a_{32}(\lambda) b_{13}(-\lambda + \alpha_1 ) &= Q^{-1} a_{14}(-\lambda + \alpha_1 ) b_{42}(\lambda) \\
a_{43}(\lambda) b_{24}(-\lambda + \alpha_2 ) &= Q^{-1} a_{21}(-\lambda + \alpha_2 ) b_{13}(\lambda) \tag{$C'$} \label{concp}
\end{align*}
\begin{align*}
a_{14}(\lambda) a_{21}(-\lambda + \alpha_2 + \alpha_3 ) &= b_{24}(\lambda) Q^{-\lambda \cdot l_2} + Q b_{24}(-\lambda + \alpha_2 + \alpha_3 ) Q^{\lambda \cdot l_4} \\
a_{21}(\lambda) a_{32}(-\lambda - \alpha_1 - \alpha_2 ) &= b_{31}(\lambda) Q^{-\lambda \cdot l_3} + Q b_{31}(-\lambda - \alpha_1 - \alpha_2 ) Q^{\lambda \cdot l_1} \\
a_{32}(\lambda) a_{43}(-\lambda - \alpha_2 - \alpha_3 ) &= b_{42}(\lambda) Q^{-\lambda \cdot l_4} + Q b_{24}(-\lambda - \alpha_2 - \alpha_3 ) Q^{\lambda \cdot l_2} \\
a_{43}(\lambda) a_{14}(-\lambda + \alpha_1 + \alpha_2 ) &= b_{13}(\lambda) Q^{-\lambda \cdot l_1} + Q b_{13}(-\lambda + \alpha_2 + \alpha_3 ) Q^{\lambda \cdot l_3}   \tag{$D'$} \; . \label{condp} 
\end{align*}
\\
there are now enough conditions to solve for $\{ a_{ij}(\lambda) \}$ and $\{ b_{ij}(\lambda) \}$ with one approach given here. note that \eqref{conbp} are the only conditions which linearly relates unknowns so \eqref{conbp} is a good place to start.

\subsection*{Solution method} \label{appa}

Examination of \eqref{conbp} reveals that 
\begin{align*}
a_{ij}(\lambda) = Q^{\frac{1}{2}\lambda \cdot (l_i + l_j)} \tilde{a}_{ij}(\lambda) 
\end{align*}
where $\tilde{a}_{ij}(\lambda) = \tilde{a}_{ij}(-\lambda - \alpha_{i-1})$, so if $\tilde{a}_{ij}$ contains a term proportional to $Q^{\lambda \cdot Y}$ then it must also contain a term proportional to $Q^{-\lambda \cdot Y}$.
\\
\\
Now consider \eqref{conap}. It would appear to be difficult to realise the likes of $b_{31}(\lambda) = \frac{1}{b_{13}(-\lambda)}$ if $b_{ij}$ contains more than one term, so a reasonable ansatz is $b_{ij}(\lambda) = B_{ij} Q^{\lambda \cdot X_{ij}}$ where $B_{ij}$ is constant. Then
\begin{align*}
b_{13}(\lambda) b_{31}(-\lambda) &= B_{13}B_{31} Q^{\lambda \cdot (X_{13} - X_{31})} = 1 \\
b_{24}(\lambda) b_{42}(-\lambda) &= B_{24}B_{42} Q^{\lambda \cdot (X_{24} - X_{42})} = 1 \; .
\end{align*}
\\
It is clear then that $X_{13} = X_{31}$ while $B_{13}B_{31}=1$, etc. A choice is made here to make the solution simple, which is that each prefactor is set to unity. The scaling of the $b_{ij}$s (and $a_{ij}$s) is investigated in appendix \ref{appb}. This choice then means that $b_{13}(\lambda) = b_{31}(\lambda)$ and $b_{24}(\lambda) = b_{42}(\lambda)$.
\\
\\
At this juncture \eqref{cone} and \eqref{cond}, with the above identifications gives $b_{ij}(\lambda + 2\alpha_k + 2\alpha_{k+1}) = b_{ij}(\lambda)$ for any $k$ modulo $4$. Given the assumptions made about the form of $b_{ij}$, The $\lambda$ dependence is entirely in the exponent of $Q$ so there is no mechanism that will give minus signs and so 
\begin{align*}
b_{ij}(\lambda + \alpha_1 + \alpha_2) &= b_{ij}(\lambda + \alpha_2 + \alpha_3) = b_{ij}(\lambda) \; .
\end{align*}
\\
A consequence of the above is then that
\begin{align*}
b_{13}(\lambda) = b_{31}(\lambda) = Q^{a \lambda \cdot (l_1 + l_3)}  & & b_{24}(\lambda) = b_{42}(\lambda) = Q^{b \lambda \cdot (l_2 + l_4)}
\end{align*}
with $a$ and $b$ constants. 
\\
\\
The knowledge that $b_{13}(\lambda) = b_{31}(\lambda)$ and $b_{24}(\lambda) = b_{42}(\lambda)$ can now be used to begin to find the form of $\tilde{a}_{ij}$. The first equation in \eqref{conb} and the first equation in \eqref{conf} combine to give $\tilde{a}_{14}(\lambda + \alpha_1 + \alpha_2) = \tilde{a}_{14}(\lambda - \alpha_1 - \alpha_2)$; similarly, the second equation in \eqref{conb} and the first in \eqref{cong} combine to give $\tilde{a}_{14}(\lambda + \alpha_2 + \alpha_3) = \tilde{a}_{14}(\lambda - \alpha_2 - \alpha_3)$. Assuming that all of the $\lambda$ dependence in $\tilde{a}_{ij}(\lambda)$ is in the exponent of powers of $Q$ the conditions of \eqref{conb}, \eqref{conf} and \eqref{cong} give
\begin{align*}
\tilde{a}_{ij}(\lambda + \alpha_1 + \alpha_2) = \tilde{a}_{ij}(\lambda + \alpha_2 + \alpha_3) = \tilde{a}_{ij}(\lambda) \; .
\end{align*}
\\
The combination of the above conditions with \eqref{conbp} implies that
\begin{align*}
\tilde{a}_{14}(\lambda) &= \tilde{a}_{14}(-\lambda - \alpha_0) = \tilde{a}_{14}(-\lambda + \alpha_1) = \tilde{a}_{14}(-\lambda - \alpha_2) = \tilde{a}_{14}(-\lambda + \alpha_3) \\
\tilde{a}_{21}(\lambda) &= \tilde{a}_{21}(-\lambda + \alpha_0) = \tilde{a}_{21}(-\lambda - \alpha_1) = \tilde{a}_{21}(-\lambda + \alpha_2) = \tilde{a}_{21}(-\lambda - \alpha_3) \\
\tilde{a}_{32}(\lambda) &= \tilde{a}_{32}(-\lambda - \alpha_0) = \tilde{a}_{32}(-\lambda + \alpha_1) = \tilde{a}_{32}(-\lambda - \alpha_2) = \tilde{a}_{32}(-\lambda + \alpha_3) \\
\tilde{a}_{43}(\lambda) &= \tilde{a}_{43}(-\lambda + \alpha_0) = \tilde{a}_{43}(-\lambda - \alpha_1) = \tilde{a}_{43}(-\lambda + \alpha_2) = \tilde{a}_{43}(-\lambda - \alpha_3) \; .
\end{align*}
\\
Since $\tilde{a}_{32}$ obeys the same conditions as $\tilde{a}_{14}$ while $\tilde{a}_{43}$ obeys the same conditions as $\tilde{a}_{21}$. It is reasonable then to make the assumption that, up to a multiplicative factor,
\begin{align*}
\tilde{a}_{14}(\lambda) = \tilde{a}_{32}(\lambda) & & \tilde{a}_{21}(\lambda) = \tilde{a}_{43}(\lambda) 
\end{align*}
something which is consistent with the conditions \eqref{conc}. With this identification $b_{ij}(\lambda)$ can now be fully determined using \eqref{concp}. The first equation of \eqref{concp} reduces to $Q^{-a \lambda \cdot (l_1 + l_3)} Q^{a} = Q^{-b \lambda \cdot (l_1 + l_3)} Q^{-\frac{1}{2}}$ and so the conclusion is that $a = b = -\frac{1}{2}$. The other terms in \eqref{concp} all agree with this identification, so 
\begin{align*}
b_{13}(\lambda) = b_{31}(\lambda) = Q^{-\frac{1}{2}\lambda \cdot (l_1 + l_3)} & & b_{24}(\lambda) = b_{42}(\lambda) = Q^{-\frac{1}{2}\lambda \cdot (l_2 + l_4)} \; .
\end{align*}
\\
The remaining conditions \eqref{cona} and \eqref{condp} can be shown to be equivalent via use of \eqref{conbp}. These inhomogeneous equations and the constraints previously found on $\tilde{a}_{ij}(\lambda)$ suggest ans\"atze of the form
\begin{align*}
\tilde{a}_{14}(\lambda) = \tilde{a}_{32}(\lambda) &= A\left(Q^{\frac{1}{2}\lambda \cdot (l_1 + l_3)} + Q^{\frac{1}{2}-\frac{1}{2}\lambda \cdot (l_1 + l_3)} \right) + B \\
\tilde{a}_{21}(\lambda) = \tilde{a}_{43}(\lambda) &= C \left(Q^{\frac{1}{2}\lambda \cdot (l_2 + l_4)} + Q^{\frac{1}{2}-\frac{1}{2}\lambda \cdot (l_2 + l_4)} \right) + D \; .
\end{align*}
\\
The first condition in \eqref{condp} becomes $\tilde{a}_{14}(\lambda) \tilde{a}_{21}(-\lambda) = Q^{-\frac{1}{2} - \lambda \cdot (l_2 + l_4)} + Q^{\frac{1}{2} + \lambda \cdot (l_2 + l_4)}$ so the above ans\"atze give
\begin{align*}
AC = Q^{-\frac{1}{2}} & & AD + BC = 0 & & BD = - 2
\end{align*}
and all of the other equations in \eqref{condp} and \eqref{cona} give the same relations. The most symmetrical solution is then to take $A = C = Q^{-\frac{1}{4}}$, so that $B = -D$ giving two solutions
\begin{align*}
\tilde{a}_{14}(\lambda) = \tilde{a}_{32}(\lambda) &=  Q^{- \frac{1}{4} + \frac{1}{2}\lambda \cdot (l_1 + l_3)} +  Q^{\frac{1}{4} -\frac{1}{2}\lambda \cdot (l_1 + l_3)} \pm \sqrt{2} \\
\tilde{a}_{21}(\lambda) = \tilde{a}_{43}(\lambda) &= Q^{- \frac{1}{4} + \frac{1}{2}\lambda \cdot (l_2 + l_4)} + Q^{\frac{1}{4} -\frac{1}{2}\lambda \cdot (l_2 + l_4)} \mp \sqrt{2} 
\end{align*}
where either the upper sign is taken for every $\tilde{a}_{ij}$ or the lower sign is taken for every $\tilde{a}_{ij}$.

\section{Rescaling and unitary transformations} \label{appb}

\subsection*{Rescaling symmetry}

Freedom to rescale the quantities $\{ a_{ij}(\lambda) \}$ and $\{ b_{ij}(\lambda) \}$ can be seen in the conditions of appendix \ref{appa} in the triangle relations \eqref{cona} - \eqref{cong} and the crossing-unitarity relations \eqref{conap} - \eqref{condp}. Labelling the scaling by $a_{ij}(\lambda) \to A_{ij}a_{ij}(\lambda)$ and $b_{ij}(\lambda) \to B_{ij}b_{ij}(\lambda)$ with $\{ A_{ij} \}$ and $\{ B_{ij} \}$ sets of constants, the solutions \eqref{sol1} and \eqref{sol2} correspond to having $A_{ij} = B_{ij} = 1$ in all cases.
\\
\\
Most of the conditions in appendix \ref{appa} do not restrict the scaling as they are homogeneous, in that both sides of the equality are scaled by the same quantities. The conditions which do restrict the scaling are \eqref{cona}, \eqref{conap}, \eqref{concp} and \eqref{condp} with both \eqref{cona} and \eqref{condp} giving
\begin{align*}
A_{14}A_{21} = B_{24} & & A_{21}A_{32} = B_{31} & & A_{32}A_{43} = B_{42} & & A_{43}A_{14} = B_{13}
\end{align*}
which is consistent with \eqref{concp}. There is just one more constraint which comes from \eqref{conap} which is that
\begin{align*}
A_{14}A_{21}A_{32}A_{43} = 1 \; .
\end{align*}
\\
Notable is that the triangle relations do not enforce the above constraint. It is clear that there are just three independent parameters that allow solutions to be rescaled consistently with both the triangle relations \eqref{triangle} and the crossing-unitarity relations \eqref{cruni}.  

\subsection*{Unitary transformations}

A further freedom in the solution comes from taking a diagonal unitary transformation $T \to UTU^{\dagger}$. In order for the triangle relations to still hold the unitary transformations considered should depend only on the defect topological charge and not on the soliton topological charge. In order to change the actual charge dependence of the $T$-matrix the unitary transformation should depend quadratically on the the defect topological charge, so such a transformation matrix has the form
\begin{align*}
U_{\alpha}^{\beta} = Q^{\tfrac{1}{2}c \alpha \cdot \alpha} \delta_{\alpha}^{\beta}
\end{align*}
with $c$ a real parameter. The effect of this unitary transformation is that the diagonal terms in the $T$-matrix are unchanged while
\begin{align*}
a_{ij}(\lambda) \to Q^c Q^{c \lambda \cdot \left( l_j - l_i \right)} a_{ij}(\lambda) \\
b_{ij}(\lambda) \to Q^c Q^{c \lambda \cdot \left( l_j - l_i \right)} b_{ij}(\lambda) \; .
\end{align*}
\\
The solution \eqref{sol1} or \eqref{sol2} is thus altered correspondingly. Note that the unitary transformation $U_{i \alpha}^{j \beta} = Q^{\frac{l_i \cdot \alpha}{4}} \delta_i^j \delta_{\alpha}^{\beta}$ considered in \cite{CZ07} and seemingly dependent on soliton labels has the same effect as the above transformation when $c = -\frac{1}{4}$.
\\
\\
A linear unitary transformation such as $U_{\alpha}^{\beta} = Q^{\alpha \cdot X} \delta_{\alpha}^{\beta}$, where $X$ is a fixed vector in the root space, gives something entirely equivalent to the rescaling symmetry already considered. Note that the general quadratic unitary transformation here does not affect the crossing-unitarity relations.

\end{document}